Sivian and White revisited: the role of resonant thermal noise pressure on the eardrum in auditory thresholds.

Michael J. Harrison
Department of Physics and Astronomy
Michigan State University
East Lansing, MI 48824

The influence of thermal pressure fluctuations on the tympanic membrane has been re-examined as a possible contributing determinant of the threshold of human hearing over the range of audible frequencies. The early approximate calculation of Sivian and White [1] is shown to result in higher values of thermal noise pressure on the tympanium of a model meatus than the result obtained by directly calculating the noise pressure from thermally excited resonant ear canal modes.

It has been known for some time that human sense organs have sensitivities that approach limits set by thermal noise, and photon quantization in the case of vision [2,3,4]. The thresholds of hearing in humans and other primates are determined by incident coherent signals whose intensity just matches the noise from thermal agitation in all parts of the auditory apparatus: the eardrum terminating the ear canal, the stapes of the middle ear, and hair cells residing on the basilar membrane of the cochlea [5,6]. But the relative importance of these three sources of thermal noise in different audio frequency intervals requires detailed discussion. In this paper we focus exclusively on the role of noise coming from thermal excitation of resonant sounds waves within the ear canal/pinna combination which results in noise pressure on the eardrum. We shall assume that such noise pressure defines the minimum auditory pressure that corresponds to the threshold of hearing. This mechanism was studied by Sivian and White [1].

Subsequent investigations have discussed the relation between thermal noise and possible active elements in the inner ear, as well as issues generally concerned with signal processing in the nervous system. Studies of quantum-mechanical effects on noise that affects hearing thresholds have also been published [7,8]. Additional. work has focused more precisely on the connection between statistical mechanics and sensory signal processing by biological means in creatures' brains [9]. Active elements within the cochlea have been identified as the underlying source of nonlinearities, which arise from regenerative signals that compensate for damping, and lead to extraordinary sharp tuning and amplification of sound [10].

The Sivian and White [1] calculation of rms noise pressure on the eardrum mimics the eardrum with a rigid massless piston of circular cross-section and area S that freely vibrates back and forth in an unflanged rigid wall exposed to the thermally fluctuating atmosphere on one side. Such a model does not capture effects arising from thermal excitation of resonant sound waves in an ear canal of finite length open at the pinna end and closed at the eardrum. This paper therefore adopts a model of the outer ear consisting of a straight tube of uniform circular cross section, open at one end and terminated at the other end by a rigid circular closure of area S that represents the eardrum [11]. We include some of the geometrical effects of the pinna flare by incorporating its extension in the total length of the straight tube, and by choosing a tube diameter somewhat larger than that of a typical average ear canal. We further stipulate a tube length that includes an addition coming from an open-end geometric length correction given by 0.6 the tube radius which represents the inertia of air set into vibration in the immediate vicinity of an unflanged open end [12].

The displacement y(x,t) of air in one-dimensional damped sound waves within the model ear canal of effective extended length L in the x-direction is governed by the wave equation [13]

$$\partial^2 y / \partial t^2 = v^2 \partial^2 y / \partial x^2 + \Gamma_0^2 [\partial^2 / \partial x^2][\partial y / \partial t] , \qquad (1)$$

where v is the speed of sound and $\Gamma_0^2$ is a damping parameter that in the case of longitudinal sound in unbounded air depends linearly on bulk and shear viscosities of air. Within a narrow ear canal that has dissipative walls, we anticipate that energy lost at the walls of the enclosure, as well as energy transmitted onward through the middle ear, may be represented by a damping parameter $\Gamma_0^2$ which we shall regard as an adjustable clinical parameter chosen to obtain agreement with observed thresholds.

For air mass density $\rho_0$ and sound speed v the pressure fluctuation $\Delta p(x,t) = -v^2 \rho_0 \partial y/\partial x$ vanishes at the open effective end at x=L, and the air displacement y(x,L) vanishes on a model rigid eardrum x=0. These boundary conditions lead to a general superposition of normal modes for time $t \geq 0$:

$$y(x,t) = \sum_n \alpha_n(t) \sin(n\pi x/2L) \quad , \tag{2}$$

where n is odd, and

$$\alpha_n(t) = A_n \, e^{-(\Gamma_0^2 \pi^2 n^2 / 8L^2)t} \, e^{-(i\omega_n t)} \tag{3}$$

is the damped normal coordinate with initial amplitude $A_n$, and $\omega_n = n\pi v / 2L$ represents the corresponding normal mode frequency. We neglect very small frequency shifts from damping. For t < 0 we take y(x,t) =0.

The total system energy consisting of kinetic and potential energy of air vibrating within the effective model ear canal is readily calculated and is given by the following Hamiltonian:

$$H = \int_0^L dx \; [S \rho_0 /2] \, [ (\partial y/\partial t)^2 + v^2 (\partial y/\partial x)^2 ] \quad , \tag{4}$$

which upon insertion of Eqn (2) for y(x,t) becomes

$$H = (S \rho_0 L/ 4) \sum_{n \, odd} | \dot{\alpha}_n |^2 + (v^2 \rho_0 S \pi^2 / 16 L) \sum_{n \, odd} | \alpha_n |^2 \, n^2 \quad , \tag{5}$$

where S is the cross-section area of the tube representing the extended effective ear canal. Since each quadratic term in the total Hamiltonian has its equipartition thermal average over a canonical ensemble we obtain

$$< |\alpha_n|^2 > = ( 8LkT / n^2 v^2 \rho_0 S \pi^2 ) \tag{6}$$

for the normal coordinates $\alpha_n$, where the brackets signify thermal averaging. At the closed end pressure antinode, x = 0, we use $\Delta p(0,t) = - v^2 \rho_0 (\partial y/\partial x|_{x=0})$ for the pressure fluctuation and with Eqn.(2) one obtains the thermal average noise pressure

$$< | \Delta p |^2 > = ( v^4 \rho_0^2 \pi^2 / 4L^2 ) \sum_{n,n'} nn' < \alpha_n \alpha_{n'}^* > \quad . \tag{7}$$

In thermal equilibrium the normal coordinates $\alpha_n$ are uncorrelated with random phases, and the double sum in Eqn.(7) reduces to a single sum over odd n :

$$<|\Delta p|^2> \; = \; (v^4 \rho_0^2 \pi^2 / 4L^2) \sum_{n \; odd} \{ n^2 <|\alpha_n|^2> \} \qquad . \qquad (8)$$

For external ear dimensions and acoustic frequencies relevant to hearing there will be a highest odd mode number N which provides a frequency cut-off in the auditory process, and in the summation above for the thermal noise pressure $<|\Delta p|^2>$ that is a match for the coherent sound pressures at hearing thresholds. From Eqn (6) and Eqn (8) we obtain

$$<|\Delta p|^2> \; \equiv \; p_{rms}^2 \; = \; [ \, ( 2v^2 \rho_0 kT / LS)(\{N+1\}/2) \, ] \qquad , \qquad (9)$$

where all odd mode frequencies up to $f_N = N(v/4L)$ contribute to the threshold noise pressure on the eardrum which leads to the auditory threshold for the included frequencies. In Eqn (9) $V = LS$ is the effective volume of air that contains thermally excited sound waves, including the volume of the pinna flare that abuts the tragus and concha. The volume V includes an effective augmented canal diameter that takes account of the large pinna flare that terminates the open end of the extended pinna – ear canal tube. And the length L additionally includes the open end correction for length given by 0.6 the tube radius of the model extended pinna – ear canal tube adopted for calculation.

We expect $f_1 = v/4L$ to be a resonant frequency located in the region of greatest auditory sensitivity. For a typical adult human ear we take effective dimensions $V = 22.5$ cm$^3$ corresponding to $L = 4.5$ cm and $S = 5.0$ cm$^2$ for effective volume and length respectively, which include the large volume of the pinna flare and the open end correction for length. For $T = 300\,^0$K, $\rho_0 = 1.18 \times 10^{-3}$ gm/cm$^3$, and $v = 3.44 \times 10^4$ cm/sec one obtains $f_1 = 1{,}911$ Hz and $p_{rms} = 1.0 \times 10^{-4}$ dynes/cm$^2$ when $N = 7$ is the highest mode number relevant to acoustically significant threshold noise. These values of $f_1$ and $p_{rms}$ are in the region of greatest observed auditory sensitivity [11]. In decibels relative to $p_0 = 2 \times 10^{-4}$ dynes/cm$^2$ this $p_{rms}$ corresponds to a threshold sound pressure of -3.0 dB.

We can use Eqn.(8) in the limits $L \rightarrow \infty$, $S \rightarrow \infty$ to derive the Sivian–White result for the mean square fluctuating pressure on the surface of an infinite reflecting rigid wall due to thermally excited sound waves in a specified frequency interval coming from the semi-infinite half-space of air facing the wall.

Rewrite Eqn.(8) for the case in which the odd integers n are restricted to be within a series sequence extending from $n_1$ to $n_2$. Then

$$<|\Delta p|^2> \; = \; (v^4 \rho_0^2 \pi^2 / 4L^2) \sum_{n_1}^{n_2} \{ n^2 <|\alpha_n|^2> \} \, . \qquad (10)$$

Recall $\omega_n = 2\pi f_n = (n\pi v / 2L)$ for the angular frequencies, so $n = (2L\omega_n / \pi v)$. Therefor $\Delta\omega_n = (\pi v/2L) \Delta n = \pi v/L$ is the separation between adjacent normal mode frequencies, since $\Delta n = 2$ for adjacent odd integers. From Eqn.(6) and Eqn.(10) we then obtain

$$<|\Delta p|^2> = v^2 \rho_0^2 \sum_{n_1}^{n_2} \omega_n^2 \ (8LkT / n^2 v^2 \rho_0 S \pi^2) \qquad . \tag{11}$$

Now let $S \to \infty$, $L \to \infty$ such that $8L / S \to 1/L$ as $L \to \infty$. The area S of the model tube cross-section and its closed end then become infinite like $S \approx 8 L^2 \to \infty$ so that the noise pressure becomes exerted on the infinite rigid plane surface considered by Sivian and White. As $L \to \infty$ and $S \to \infty$ we then obtain the asymptotic value

$$( 8LkT / n^2 v^2 \rho_0 S \pi^2 ) \to ( kT / n^2 v^2 \rho_0 L \pi^2 ) \qquad . \tag{12}$$

For large L we define the differential quantity $d\omega_n \equiv v / n^2 L$. Then $d\omega_n \to 0$ as $n^2 L \to \infty$ for any finite n. Label the thermal averages $<|\alpha_n|^2>$ according to associated normal mode frequencies obtained from Eqn.(6), and obtain:

$$< |\alpha_n|^2 > \equiv ( kT / v^3 \rho_0 \pi^2 ) d\omega_n \qquad . \tag{13}$$

The thermal noise pressure on the rigid wall at $x = 0$ coming from frequencies included in the interval from $\omega_{n1}$ to $\omega_{n2}$ is then

$$<|\Delta p|^2> = ( \rho_0 kT / v\pi^2 ) \sum_{\omega_{n1}}^{\omega_{n2}} \omega_n^2 \ d\omega_n \tag{14}$$

in the asymptotic region $L \to \infty$ and $S \to \infty$ as $8L^2 \to \infty$.
As $L \to \infty$ Eqn.(14) passes from a sum to an integral:

$$<|\Delta p|^2> = ( \rho_0 kT / v\pi^2 ) \int_{\omega_1}^{\omega_2} \omega^2 \ d\omega \qquad . \tag{15}$$

Performing the integration in Eqn.(15) and recalling that $\omega = 2\pi f$ we obtain

$$<|\Delta p|^2> = ( 8\rho_0 \pi kT / 3v ) [ f_2^3 - f_1^3 ] \qquad , \tag{16}$$

which is exactly the result obtained by Sivian and White [1].

We calculate the rms pressure $p_{rms} = \sqrt{\langle |\Delta p|^2 \rangle}$ using Eqn.(16) for the same interval $f_1 = 1,911$ Hz and $f_2 = 7f_1$ adopted to evaluate Eqn.(9) for the stipulated finite ear-canal to pinna flare geometry. For the same air density, temperature, and sound speed we now obtain from Eqn.(16) $p_{rms} = \sqrt{\langle |\Delta p|^2 \rangle} = 2 \times 10^{-4}$ dynes/cm$^2$ for the Sivian-White result. The noise pressure on a model eardrum according to Sivian-White is 3 decibels higher than that obtained from noise pressure contributions from resonant thermally excited ear canal modes. We introduce the neurological ansatz that the auditory threshold sensitivity at any frequency, measured by threshold pressure on the eardrum, is given primarily by the noise pressure incident on the eardrum from thermal excitations developed in the ear canal. Other sources of noise from random fluctuations of middle ear stapes or random motions of hair cells connected to the basilar membrane in the cochlea are of secondary importance in creating noise in the sensorineural system that limits clear signal propagation to the brain and thereby helps create auditory thresholds.

The model studied in this paper permits us to conclude that auditory threshold sensitivity in the frequency interval from 1,911 Hz to 13,371 Hz is likely to be several decibels more acute than the Sivian and White result [1] would indicate.